\documentclass{iaus}
\usepackage{graphicx}

\title[Young open clusters in the MW and SMC] 
{Young open clusters in the Milky Way and Small Magellanic Cloud}

\author[C. Martayan]   
{C. Martayan$^{1,2}$}

\affiliation{$^1$ European Organisation for Astronomical Research in the Southern 
Hemisphere, Alonso de Cordova 3107, Vitacura, Casilla 19001, Santiago 19, Chile \\ email: {\tt cmartaya@eso.org} \\[\affilskip]
$^2$ GEPI, Observatoire de Paris, CNRS, Universit\'e Paris Diderot, 
5 place Jules Janssen, 92195 Meudon Cedex, France}

\pubyear{2009}
\volume{266}  
\pagerange{119--126}
\jname{Star Clusters}
\editors{Richard de Grijs \& Jacques R. D. L\'epine, eds.}
\begin{document}

\maketitle

\begin{abstract}
NGC6611, Trumpler 14, Trumpler 15, Trumpler 16, Collinder 232 are very young open clusters
located in star-formation regions of the Eagle Nebula or the Carina in the MW,
and NGC346 in the SMC. With different instrumentations and techniques, it was
possible to detect and classify new Herbig Ae/Be stars, classical Be stars and
to provide new tests / comparisons about the Be stars appearance models. Special
stars (He-strong) of these star-formation regions are also presented. 
\keywords{open clusters and associations: individual (Trumpler 14, Trumpler 15, Trumpler
16, Collinder 232), Magellanic Clouds, galaxies: star clusters (NGC346), stars: early-type,
stars: pre--main-sequence, stars: emission-line, Be, stars: evolution, ISM: dust, extinction}
\end{abstract}

\firstsection 
\section{Observations}

We used the ESO-WFI (\cite[Baade et al. 1999]{Baade et al. 1999}) in its slitless mode (in
Halpha, R$\sim$1000) for observing the Galactic open cluster NGC6611, which lies in
the Eagle Nebula. This instrumentation allows to disentangle
the star with emission-lines from a true circumstellar disk than nebular lines
from the diffuse emission of the surrounding nebula. We then observed 100 stars
of NGC6611 pre-selected from our WFI catalogue with the VLT-FLAMES/GIRAFFE 
facilities in MEDUSA mode (\cite[Pasquini et al. 2002]{Pasquini et al. 2002}). About the open
clusters Trumpler 14, 15, 16, and Collinder 232, which lie in the Carina region, 
we pre-selected the B-type stars from our astrophotometric catalogues
based on the EIS pre-FLAMES survey images (\cite[Momany et al. 2001]{Momany et al. 2001}). We then
observed $\sim$200 of OB-type stars with the VLT-FLAMES/GIRAFFE. We used the LR2, HR4, HR5,
HR6, and HR15 settings with resolutions from 6500 to 18000. These setups were
chosen for the Balmer lines, Helium I and II lines, Mg II lines, Si III lines. 
The SMC open cluster NGC346 was observed with the ESO-WFI (\cite[Baade et al. 1999]{Baade et al. 1999}) 
in its slitless mode (in Halpha). More than 50 emission-line stars were
found, half of them were detected for the first time, see \cite[Martayan et al. (2009)]{Martayan et al. 2009}.

\section{Determination of the interstellar reddening and correction of the magnitudes}

\begin{figure}[b]
\begin{center}
 \includegraphics[width=3.4in, angle=0]{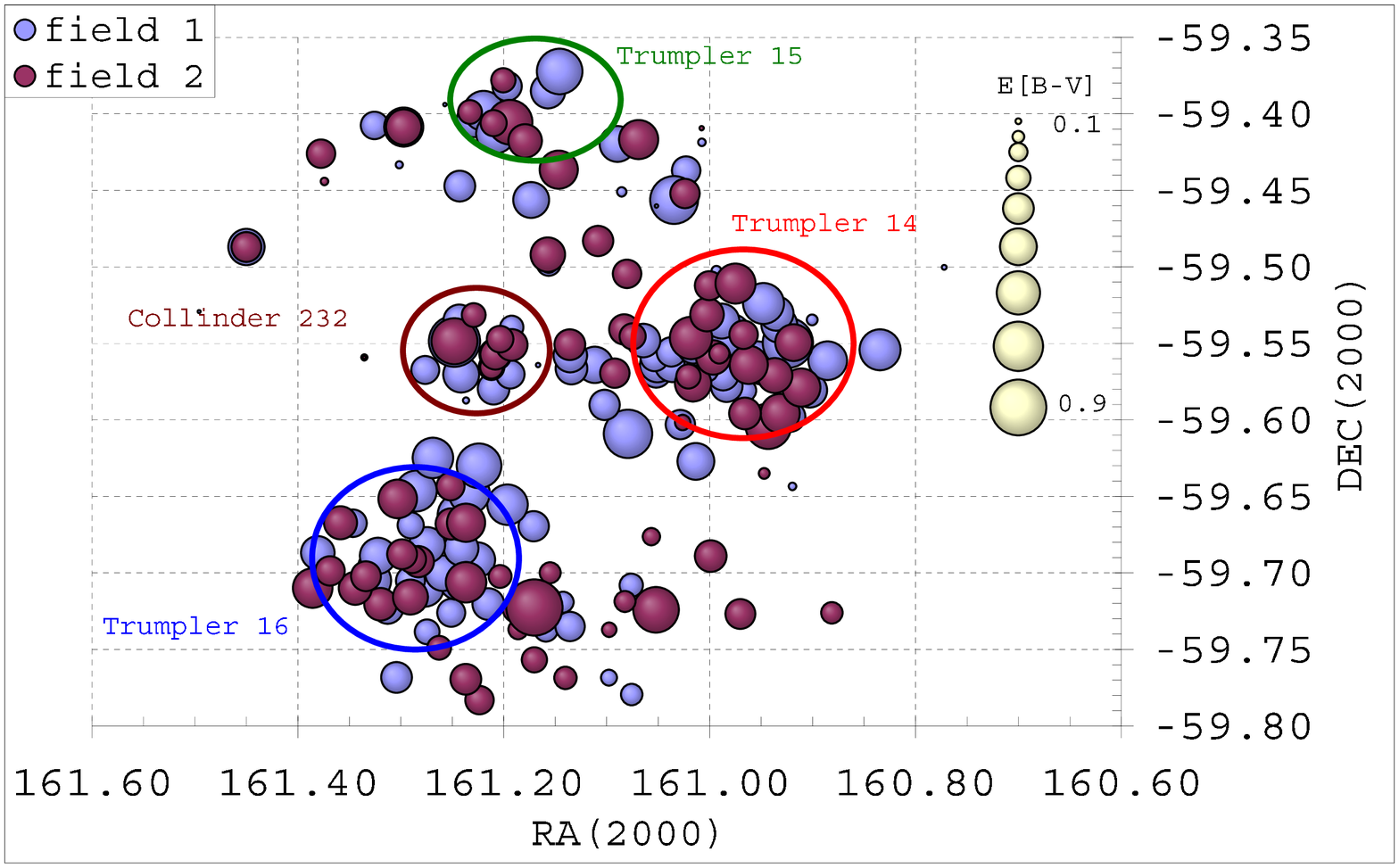} 
 \caption{Map of the interstellar reddening in the Carina region for stars observed in Trumpler 14, 15, 16, and Collinder 232.}
   \label{fig1p577}
\end{center}
\end{figure}

\begin{figure}[b]
\begin{center}
 \includegraphics[width=3.4in, angle=0]{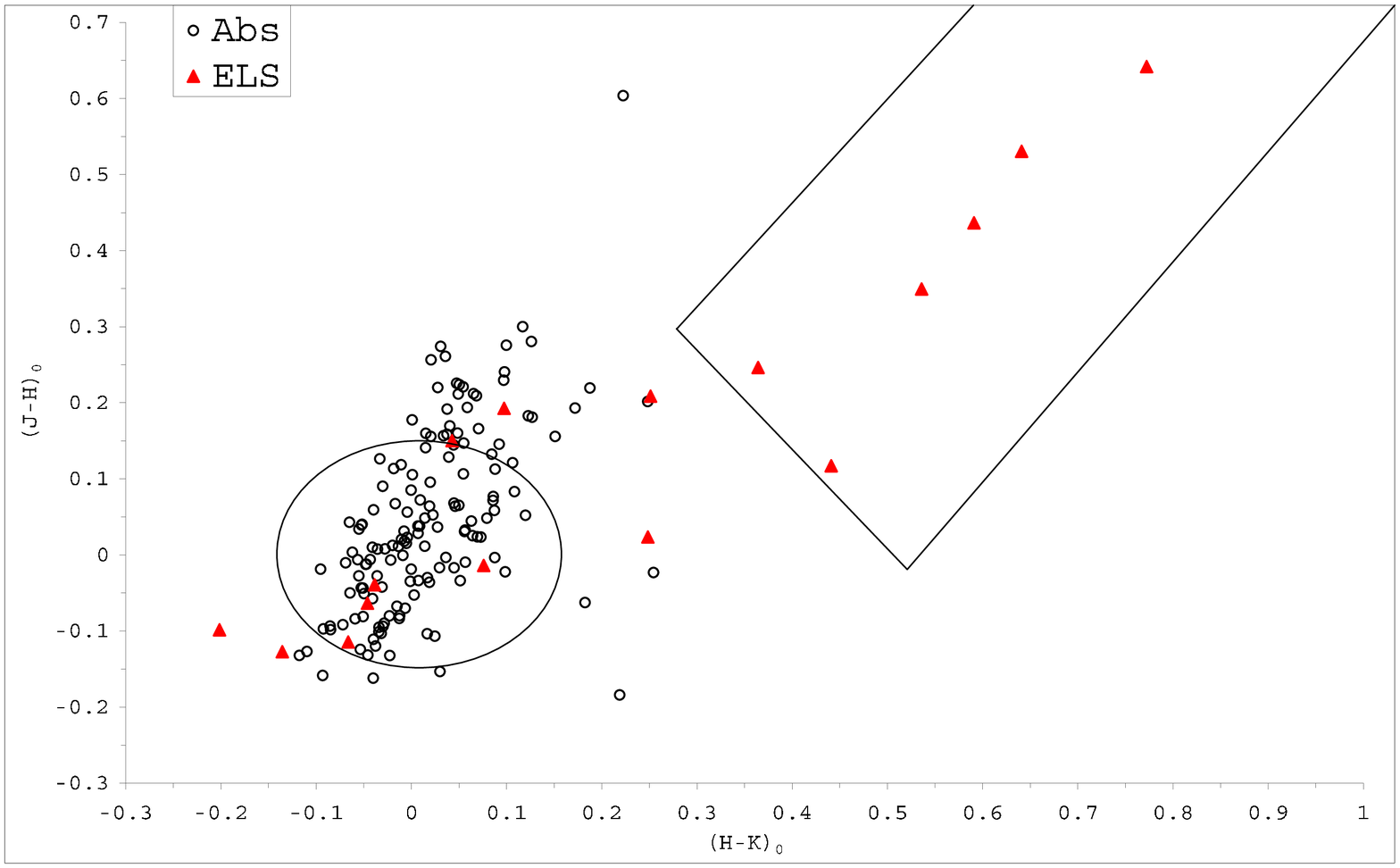} 
 \caption{Dereddened 2MASS colour-colour diagram for stars observed in Trumpler 14, 15, 16, Col232 open clusters.
 The large rectangular area from \cite[Hern{\'a}ndez et al. (2005)]{hernandez05} corresponds to the area of PMS objects like Herbig Ae/Be stars.}
   \label{fig2p577}
\end{center}
\end{figure}

 To correct the magnitudes of the stars of the interstellar reddening, we
measured the equivalent width of interstellar lines at 443.0, 450.2, and 661.3
nm. We then used the calibration from \cite[Herbig (1975)]{Herbig (1975)} to obtain the values of the
interstellar reddening E[B-V] for each star observed with GIRAFFE.  The map of
the reddening for the Trumpler 14, 15, 16 clusters is shown in Fig.~\ref{fig1p577}. From this
estimate of the E[B-V], we corrected the JHK magnitudes from 2MASS as well as
the SPITZER magnitudes. However, the SPITZER data are available for the stars in NGC6611 but not for the stars in 
Trumpler 14, 15, 16, open clusters. Then 2 kinds of dereddened colour-magnitudes
diagrams are obtained for NGC6611 (see \cite[Martayan et al. 2008]{Martayan et al. (2008)})
For the Trumpler 14, 15, 16 clusters, the 2MASS-diagram is shown in Fig.~\ref{fig2p577}. 
The Emission-line stars are indicated by a red triangle. 
The status of several of them is obvious: clearly PMS or Ms object.
However, for the stars located in the intermediate regions, a
complementary analysis must be done by using the evolutionary status determined
from the fundamental parameters and compared to the age of the open clusters.

\section{Fundamental parameters determination}
 
The fundamental parameters were determined by using the GIRFIT code from \cite[Fr\'emat et al. (2006)]{Fremat et
al. 2006}. This code fits the observed spectrum with theoretical ones
obtained with TLUSTY and SYNSPEC codes from \cite[Hubeny \& Lanz (1995)]{Hubeny et Lanz (1995)}. For each star
member of the open cluster (NGC6611, Trumpler 14, etc), we have obtained its age
with theoretical evolutionary tracks from \cite[Schaller et al. (1992)]{Schaller et al. (1992)}. Then the age
of the open cluster is redetermined using these stars, members of the clusters.

\section{Nature of the emission-line stars}
 
 In order to determine the exact nature of emission-line objects in NGC6611,
Trumpler 14, Trumpler 15, Trumpler 16, Collinder 232, we combined the different
parameters from the colour-colour diagrams and from the evolutionary status
determined via the HR evolutionary tracks.\\

-if the star is located in the ``box'' of Herbig Ae/Be star from \cite[Hern{\'a}ndez et al. (2005)]{hernandez05} in the dereddened 2MASS
diagram, then the star is a pre-MS star.  \\
-if the star is located in the class II or class I areas in the SPITZER diagram,
then the star is a pre-MS star.\\
-if the star is located in an intermediate region, the evolutionary status
obtained from the fundamental parameters combined to the membership of open
clusters must be taken into account:\\
	* star member of the open cluster, with an age equivalent to the open cluster $=>$ MS star\\
	* star member of the open cluster, with an age apparently older than  the cluster $=>$ pre-MS star\\
	* star not member of the open cluster $=>$ status not certain\\
-if the star is not present in the boxes of pre-MS star and its evolutionary
status compatible with the age of its hosting open cluster, then the star is
probably a classical Be star (with a decretion disk vs. an accretion disk in
case of pre-MS star like Herbig Ae/Be stars).
All the details about NGC6611 can be retrieved from \cite[Martayan et al. (2008)]{Martayan et al. (2008)}.

\section{Special stars}
Among the emission-line stars of NGC6611, 2 are of special interests, the first
one is W601, which was found to host a magnetic field by \cite[Alecian et al. (2008)]{Alecian et al. (2008)}.
They also found that this star is also an He-strong star.  Thanks to VLT-GIRAFFE
observations of W601 and W080, it was possible to compare the spectral features
of these 2 stars. We found that W080 is like a spectral twin of W601 and then is
probably also an He-strong star. Moreover, its broad, deep spectral lines cannot
be reproduced by the models (\cite[Martayan et al. 2008]{Martayan et al. (2008)}). 
In order to find if  W080 hosts a magnetic field
like W601, two sets of observations were performed at the CFHT with the ESPADONS spectropolarimeter.
However, the faintness of W080 implies that this star is at the limit of the
capabilities of this instrumentation. The preliminary results seem to show no
strong magnetic field in W080, however, the data are currently analysed to find
a potential weaker magnetic field and/or to provide the upper estimates of its
intensity as well as the confirmation of its He-strong status.

\section{Conclusions}
 We found 11 emission-line star in NGC6611 and its vicinity, 9 of them are new
Herbig Ae/Be stars.   In the open clusters Trumpler 14, 15, 16, Collinder 232
and their vicinity in the Carina nebula; 6 of the 16 emission-line stars seem to
be  Herbig Ae/Be stars. For the 10 of the remaining emission-line stars: 
the status is uncertain for 2 of them, and 8 of them are probably main sequence stars
so are classical Be stars. In NGC346, complementary observations at higher
spectral resolution are needed for determining the exact nature of the 50
emission-line stars found.

\end{document}